\documentclass[aps,prl,twocolumn,superscriptaddress,showkeys]{revtex4-1}

\usepackage{amsmath,amssymb,amsfonts}
\usepackage{graphicx}
\usepackage[dvipsnames]{xcolor}
\usepackage{comment}
\usepackage[overload]{empheq}
\usepackage{upgreek}
\usepackage{lmodern,pdftexcmds}
\usepackage[sort&compress]{natbib}
\usepackage{hyperref}
\usepackage{soul}

\graphicspath{{figures/}}

\hypersetup{
    bookmarks=true,         % show bookmarks bar?
    bookmarksopen=true,
    unicode=false,          % non-Latin characters in Acrobat’s bookmarks
    pdftoolbar=true,        % show Acrobat’s toolbar?
    pdfmenubar=true,        % show Acrobat’s menu?
    pdffitwindow=false,     % window fit to page when opened
    pdfstartview={Fit},    % fits the width of the page to the window
    pdftitle={},    % title
    pdfauthor={},     % author
    colorlinks=true,       % false: boxed links; true: colored links
    linkcolor=black,          % color of internal links
    citecolor=black,        % color of links to bibliography
    urlcolor=black,           % color of external links
    pdfborder={0 0 0}
}

\renewcommand{\bm}[1]{\boldsymbol{\mathbf{#1}}}

\providecommand*{\ppi[3][]}{\partial_{#3}^{#1}#2}
\newcommand{\qmbox}[1]{\quad\mbox{#1}\quad}

\providecommand*{\rme}{\mathrm{e}}
\providecommand*{\rmd}{\mathrm{d}}
\providecommand*{\rmi}{\mathrm{i}}
\providecommand*{\rmK}{\mathrm{K}}

\providecommand*{\ez}{\bm{e}_z}
\providecommand*{\er}{\bm{e}_{r}}

\providecommand*{\ey}{\bm{e}_{y}}

\providecommand*{\etheta}{\bm{e}_{\theta}}

\begin{document}

\preprint{}

\title{Surfing its own wave: hydroelasticity of a particle near a membrane}

\author{Bhargav Rallabandi}
%\email{vbr@princeton.edu}
\affiliation{Department of Mechanical and Aerospace Engineering, Princeton University, Princeton, New Jersey 08544, USA}

\author{Naomi Oppenheimer}
%\email{naomiop@gmail.com}
\affiliation{Center of Computational Biology, Flatiron Institute, Simons Foundation, New York, NY 10010, USA}

\author{Matan Yah Ben Zion}
%\email{matanbz@gmail.com}
\affiliation{Center for Soft Matter Research, New York University, NY 10003, USA}

\author{Howard A. Stone}
\email{hastone@princeton.edu}
\affiliation{Department of Mechanical and Aerospace Engineering, Princeton University, Princeton, New Jersey 08544, USA}

\date{\today}

\begin{abstract}
We show using theory and experiments that a small particle moving along an elastic membrane through a viscous fluid is repelled from the membrane due to hydro-elastic forces. The viscous stress field produces an elastic disturbance leading to particle-wave coupling. We derive an analytic expression for the particle trajectory in the lubrication limit, bypassing the construction of the detailed velocity and pressure fields. The normal force is quadratic in the parallel speed, and is a function of the tension and bending resistance of the membrane. Experimentally, we measure the normal displacement of spheres sedimenting along an elastic membrane and find quantitative agreement with the theoretical predictions with no fitting parameters. We experimentally demonstrate the effect to be strong enough for particle separation and sorting. We discuss the significance of these results for bio-membranes and propose our model for membrane elasticity measurements.

\end{abstract}

% insert suggested PACS numbers in braces on next line
% insert suggested keywords - APS authors don't need to do this

\maketitle

Low Reynolds number hydrodynamics prohibits a net normal force on a spherical particle moving along a rigid impermeable wall. Repulsion or attraction of the particle violates time reversal symmetry. However, relaxing the constraint of rigidity breaks %wall's rigidity constraint, and allowing it to deform, 
this symmetry; a rigid sphere moving along a soft wall (or a soft sphere along a rigid wall \cite{abk02}) experiences a repulsive force. Such a force has been shown to reduce friction near compressed and sheared elastic substrates \citep{coy88_rollcoating,sek93, sko04,sko05,sno13, sai16}. Here, we show that for a thin membrane, the effect can be orders of magnitude greater, leading to sizable displacement of suspended particles. Interactions between cell membranes and surfaces are common in many physiological and cellular processes, including blood flow in capillaries \cite{sec86, nog05, dzw03} and filtration in the spleen \cite{jan56, piv16_spleenbiomechanics}, endocytosis \cite{gol979}, and micro-swimming near interfaces \cite{tro08_softswimming, gia2010, dia13_swimmingnearmembrane, led13_microswimmer_elasticconfinement, bor13_swimming_spermatozoa_surfaces, lus2014}. Understanding these interactions on the nano and micro scales is important for efficient drug delivery and release as they significantly modify the hydrodynamic mobilities of particles such as proteins \cite{lod1995, fra2003, kim2005}. Recent work has quantified many aspects of particle hydrodynamics near membranes \cite{bic06_brownian, bic07_mobility, dad17_axisym, dad17_sphericalmembrane}, but has not addressed nonlinear interactions producing repulsive forces. 

\begin{figure}[t!]
\centering
\includegraphics[scale=1]{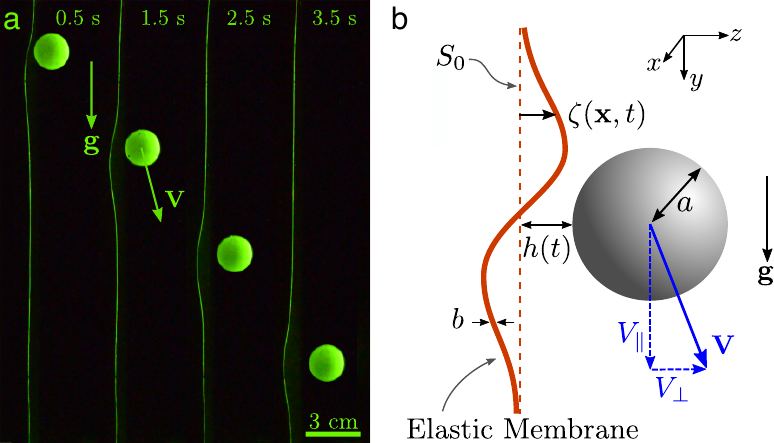}
\caption{Self surfing. (a) Time sequence showing experimental images of a solid sphere sedimenting under gravity near a vertically suspended rubber sheet in silicone oil. Tracking the position of the sphere at different times after release shows a spontaneous migration away from the sheet. (b) Sketch of the system indicating the coordinate system and relevant parameters. The dashed line indicates the undeformed position $S_0$ $(z=0)$ of the elastic membrane.}
\label{FigSetup}
\end{figure}

\begin{figure*}[t!]
\centering
\includegraphics[scale=1]{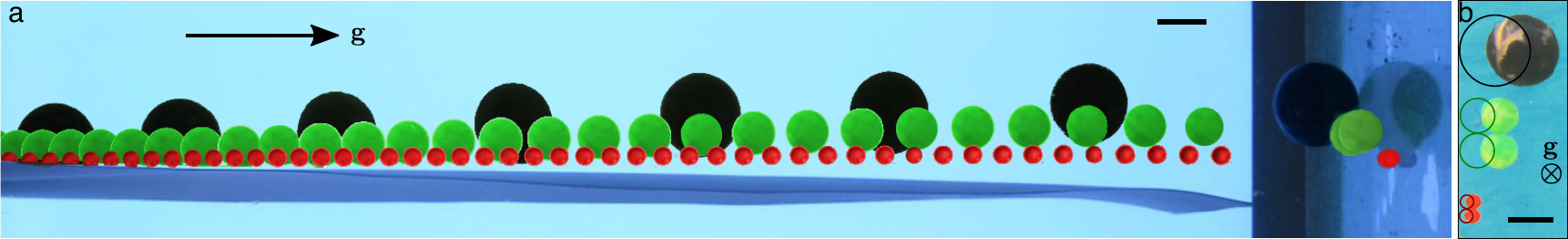}
\caption{(a) Overlaid stroboscopic images indicating the motion of three different particles [all Delrin, with radii 1.5 mm (red), 4 mm (green), 8 mm (black)], and (b) showing their rest positions on the tank floor. The particles are sorted by their properties by the end of their fall. Final rest positions of the particles as viewed from the side (a) or from above (b), showing that the deflection depends on size.  All particles in are released at the top of the tank with the same initial surface-to-surface separation distance from the sheet; initial positions are indicated as circular outlines in (b). The scale bar is 1 cm in both panels.}
\label{FigSorting}
\end{figure*}

In this Letter, we develop analytic theory and model experiments to show that a suspended particle translating tangent to a thin elastic membrane (velocity $V_\parallel$) through a viscous fluid experiences a significant migration away from the surface (velocity $V_\perp$) as a result of fluid-mediated deformations of the membrane. Figure~\ref{FigSetup}(a) shows snapshots, at different times, of a sphere sedimenting (due to gravity) along the surface of an elastic sheet that is vertically suspended in a liquid bath (experimental details provided below). The sedimentation of the sphere is accompanied by a traveling-wave deformation of the membrane [$z = \zeta(\bm{x},t)$; see Fig. \ref{FigSetup}(a,b)] and a migration of the particle away from it, resembling a particle surfing its own wave \citep{bush2015_pilot}. We find that the normal motion is sensitive to the particle size and is strong enough for particle separation and sorting; see Fig.~\ref{FigSorting}. We show that the repulsive migration velocity for a thin membrane can be several orders of magnitude greater than that obtained for a bulk elastic solid.

The magnitude of the normal velocity depends on the particles' size and density. Figure \ref{FigSorting}(a) shows a stroboscopic image of three different-sized particles (taken at fixed intervals of 0.33 s) sedimenting along the elastic sheet. The particles accelerate in the direction of gravity as a consequence of their increasing distance from the sheet over time (resulting in decreasing drag), which appears as an increasing separation distance between consecutive frames. 
Once settled, the particles are separated and sorted by their size (see Fig.~2b). Larger and heavier particles experience stronger repulsion, settling further from the sheet. These results suggest the possibility of designing continuous-throughput size-sorting devices by incorporating flexible structures in fluidic systems. 

We develop a theoretical framework to determine the repulsive velocity $V_{\perp}$ of the particle, accounting for hydrodynamic interactions that lead to a small (but finite) deformation of a nearby elastic membrane. The membrane has a bending rigidity $B$ and is held taut under a tension (force per length) $T$, both of which keep it planar in its undeformed state $S_{0}$ $(z=0)$. We note that in our experiments, the tension is a consequence of the sheet being suspended under its own weight. The advective nature of the membrane deformation makes it convenient to describe the fluid-elastic problem in a cylindrical coordinate system $\bm{r} = (r,\theta, z)$ attached to the sphere, with its origin at the point on $S_{0}$ closest to the sphere (Fig. \ref{FigSetup}b). The sphere translates with velocity $\bm{V} = V_{\parallel} \ey + V_{\perp} \ez$ with a separation distance $h(t)$ that increases in time (see Fig. \ref{FigSetup}), exciting a fluid flow (velocity $\bm{v}$ and stress $\bm{\sigma}$) that act to deform the membrane to a new position $z = \zeta(\bm{r}, t)$. 

We consider the limit of small deflections $|\zeta| \ll h$, small separation distances $h \ll a$, and predominantly parallel motion  $(V_{\parallel} \gg V_{\perp})$. In this case, the normal stress on the membrane is dominated by the fluid pressure. For small deformations of the membrane, this pressure is approximately $p^{(0)} = \frac{6 \mu V_{\parallel} \ell R \cos \theta}{5 h^2 (1+R^2)^2}$, produced by the translation of a sphere parallel to the plane $S_{0}$ \citep{one67_lubrication_sphereparallel}. Here, $\ell = \sqrt{2 a h}$ is the characteristic length scale over which stresses decay away from the sphere, and $R = r/\ell$. The membrane deformation $\zeta$ is established by a balance of this pressure with elastic stresses, $p^{(0)} = -(B \nabla^4 - T \nabla^2)\zeta$, where $\nabla$ is the 2D gradient in the $(r,\theta)$ plane \citep{landau_lifschitz86_elasticity, hel1973, sei1997}. Defining $\bm{R} = \bm{r}/\ell$, the deformation in Fourier space ($\hat{f}(\bm{k}) = \int_{\mathbb{R}^2}f(\bm{R}) \rm^{-\rmi \bm{k} \cdot \bm{R}} \rmd^2 \bm{R}$) is
\begin{subequations} 
\begin{align}
&\hat{\zeta}(\bm{k}) =    \frac{6 \pi \rmi \Lambda k a H }{5 \left(k^4 + \tau H k^2\right)} \,   \rmK_{0}(k) \cos \varphi, \label{zetahat} \quad \qmbox{with} \\    
& \Lambda = \frac{2^{\frac{5}{2}}\mu V_{\parallel}}{B}\left(\frac{a}{h}\right)^{\frac{1}{2}}, \quad \tau = \frac{2 T a^2}{B} \qmbox{and} H = \frac{h}{a} \label{LambdaBetaH},
\end{align}
\end{subequations}
where $k = |\bm{k}|$, $\varphi$ is the associated polar angle, $\rmK_0$ is the order-zero modified Bessel function of the second kind, $\Lambda = O(|\zeta|/h)$ is the (small)  deformation amplitude relative to the gap height, $\tau$ is a dimensionless tension, and $H$ is the dimensionless separation distance, which varies in time.

Next, we calculate the resulting normal velocity of the sphere, $V_{\perp}$, which arises due to a perturbation of the lubrication pressure (i.e. $p = p^{(0)} + \Lambda p^{(1)} + \cdots$). We bypass the need to solve the hydrodynamic-elastic problem at $O(\Lambda)$ (and explicitly calculate $p^{(1)}$) by using the Lorentz reciprocal theorem for viscous flows \citep{happel_book}. Introducing the known fluid velocity and stress fields ($\bm{v}'$ and $\bm{\sigma}'$) around a sphere moving perpendicular to a \emph{rigid} wall, the reciprocal statement is $\int_{S} \bm{n} \cdot \bm{\sigma} \cdot \bm{v}' \rmd^2\bm{r} = \int_{S} \bm{n} \cdot \bm{\sigma}' \cdot \bm{v}\, \rmd^2\bm{r}$, where $\bm{n}$ is the unit normal to the surface. The integration is over the \emph{undeformed} bounding surface of the fluid domain $S$, which comprises $S_{0}$, the particle surface $S_{p}$ and a surface at infinity $S_{\infty}$. With the conditions $\bm{v}  = \bm{v}' = \bm{0}$ on $S_p$ (no-slip), and $\bm{v} = -\bm{V}$ (and $\bm{v}' = -\bm{V}'$) on $S_{\infty}$, the reciprocal relation becomes $
    F_{\perp}' V_{\perp} = \int_{S_{0}}\!\!\! \bm{n} \cdot \bm{\sigma}' \cdot \left(\bm{v} + \bm{V}\right) \rmd^2 \bm{r} $, 
where $F_{\perp}' = 6 \pi \mu a^2 V'/h$ is the applied force on the sphere in the auxiliary (primed) problem \citep{cox67}.

To obtain $\bm{v}$ on $S_{0}$ $(z=0)$, we use the no-slip condition $\bm{v}|_{z = \zeta} = -\bm{V} - \ez V_{\parallel} \ppi{\zeta}{y}$, where we have assumed a quasi-static membrane deformation in the particle reference frame $(|\ppi{\zeta}{t}| \ll V_{\parallel}|\ppi{\zeta}{y}|)$. A Taylor expansion about the undeformed state then yields $\bm{v}|_{S_{0}}  \approx - \bm{V} - \ez V_{\parallel} \ppi{\zeta}{y} - \zeta \ppi{\bm{v}}{z}|_{S_{0}}$. The final step is to approximate the velocity gradient by that of the zeroth-order problem (translation parallel to a planar wall), $\ppi{\bm{v}}{z} \approx \ppi{\bm{v}^{(0)}}{z}$, leading to 
\begin{align}  \label{VnormalRT}
  V_{\perp}  &= \frac{h}{6 \pi \mu a^2 V'} \int_{S_{0}}\!\!\!\! \left( p' V_{\parallel}  \ppi{}{y} - \mu \ppi{\bm{v}'}{z} \cdot \ppi{\bm{v}^{(0)}}{z} \right) \zeta \,  \rmd^2 \bm{r}.
\end{align}

\begin{figure}[t!]
\centering
\includegraphics[scale=1]{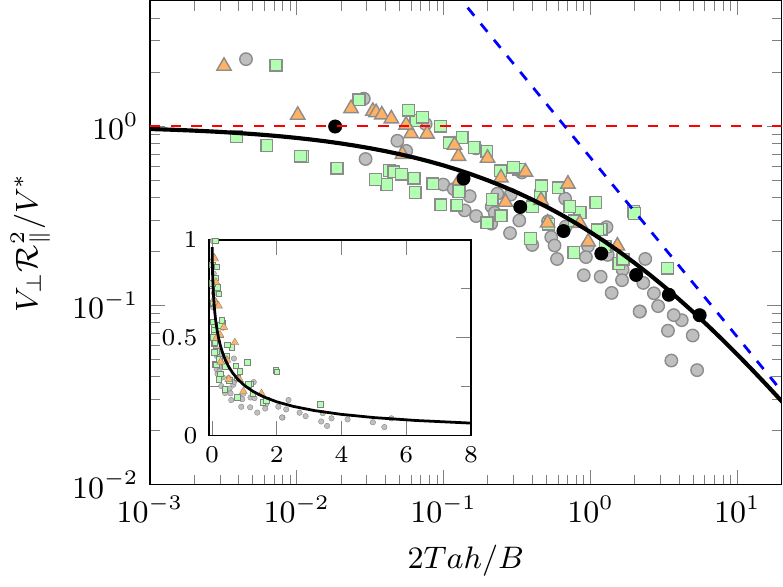}
\caption{Rescaled normal velocity $V_{\perp} \mathcal{R}_{\parallel}^2(H)/V^*$ versus the dimensionless tension $\tau H  = \frac{2 T a h}{B}$. The theoretical prediction for this quantity is the function $\mathcal{F}(\tau H)$ [see \eqref{VperpGrav}], shown as a solid black curve with asymptotes indicated as dashed lines. Experimental data for several parameter combinations (circles: 0.25 mm; squares: 0.38 mm; triangles: 0.5 mm thick sheets) are in agreement with the theoretical prediction (inset shows the same data on a linear graph). Each set of symbols shows data corresponding to several different sphere radii and densities. Since $h$ increases in time, each sphere samples a range of $2Tah/B$ values over its trajectory. The black circles indicate one such trajectory (glass sphere, $a = 5$mm; $b$ = 0.25 mm), where the value of $2Tah/B$ is initially small (due to small $h$; bending dominated) and decreases over the course of the motion (tension dominated).}
\label{FigF}
\end{figure}

We evaluate the integral \eqref{VnormalRT} in Fourier space by applying Parseval's identity and using known results for $p'$, $\ppi{\bm{v}'}{z}$ and $\ppi{\bm{v}^{(0)}}{z}$ from standard lubrication theory \footnote{$
p' = -\frac{3\mu V' \ell^2}{2 h^3 (1+R^2)^2}$, $\ppi{\bm{v}'}{z}|_{z=0} = -\frac{3 V' \ell R}{h^2 (1+R^2)^2} \er$, and $\ppi{\bm{v}^{\left(0\right)}}{z}\big|_{z=0} = \frac{2 V_{\parallel}}{5 h (1+R^2)} \left\{ \left(7 - \frac{6}{1 + R^2}\right) \er\cos \theta  - 2 \etheta \sin \theta  \right\}$}. Utilizing the expression \eqref{zetahat} for $\hat{\zeta}(\bm{k})$, we obtain
\begin{subequations} \label{Vperp}
\begin{align}
    V_{\perp} &= \frac{3 \mu a^2 V_{\parallel}^2}{25 B}\, \mathcal{F}(\tau H), \qmbox{where} \\
      \mathcal{F}\left(\tau H \right) &=\int_0^{\infty}  \frac{2 k^4 \rmK_0^2(k)}{k^4 + \tau H k^2} k \rmd k.
\end{align}
\end{subequations}
The normal drift velocity is independent of the mechanism driving tangential motion along the membrane, which can be external (gravity, magnetic field), or internal (self propulsion). The positive-definite function $\mathcal{F}(\tau H)$ decays monotonically from unity ($\tau H\ll 1$; bending dominates) to $\frac{2}{3 \tau H}$ ($\tau H \gg 1$, tension dominates) as shown in Fig. \ref{FigF}; we note that $\mathcal{F}$ can be alternatively expressed in terms of special functions. Thus, the sphere experiences a repulsive normal velocity $V_{\perp}$ that is quadratic in its speed $V_{\parallel}$ along the membrane. This quadratic dependence breaks kinematic reversibility, i.e. the sphere migrates away from the sheet irrespective of the direction of its tangential motion. We note that this result is consistent with previous studies reporting lift forces near soft substrates \citep{sek93,sko04}, albeit is here several orders of magnitude greater.

In our experiments, the driving force is gravity, which is balanced by a viscous drag to establish $V_{\parallel} = 2 a^2 g (\Delta \rho_p)/( 9 \mu \mathcal{R}_{\parallel})$, where $\Delta \rho_p = \rho_p - \rho_f$. The dimensionless resistance to tangential motion, $\mathcal{R}_{\parallel}\!\left(H\right)$, can be approximated by its limiting form for translation along a rigid plane, provided in \cite{gol67a}. Substituting the above expression for $V_{\parallel}$ into (\ref{Vperp}a) yields
\begin{equation} \label{VperpGrav}
    V_{\perp} =  V^* \frac{\mathcal{F}(\tau H)}{\mathcal{R}_{\parallel}^2(H)} \qmbox{with} V^* = \frac{4 a^6 g^2 (\Delta \rho_p)^2 }{675 \mu B}.
\end{equation}
The relative importance of tension to bending resistance in $V_{\perp}$ is quantified by the dimensionless parameter $\tau H = \frac{2Tah}{B}$. Since $H$ increases in time, this ratio is not constant during the motion of a particle. Either tension or bending may dominate during different parts of the trajectory. 

\begin{figure}[h!]
\centering
\includegraphics[scale=1]{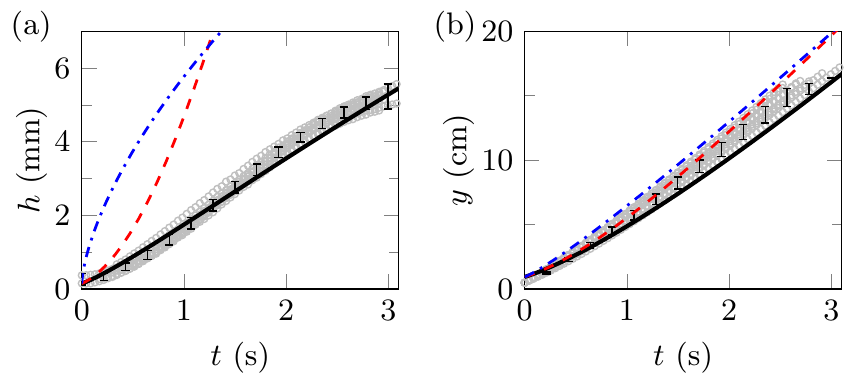}
\caption{Typical trajectories of a sphere, indicating (a) normal displacement $h(t)$ and (b) tangential displacement $y(t)$. Symbols are experimental measurements (error bars indicate one standard deviation), and the solid line corresponds to the theoretical prediction accounting for bending and tension. Bending-dominated (red dashed) and tension-dominated (blue dash-dotted) limits are indicated for comparison.}
\label{FigSampleTraj}
\end{figure}

Our experimental setup consists of silicone rubber sheets (8 cm $\times$ length 30 cm; density $\rho_s = 1.1$ g/cm$^{3}$) of 
thicknesses $b$ (0.25, 0.38, 0.5 mm) that are suspended by gravity in silicone oil (density $0.97$ g/cm$^{3}$, viscosity $1$ Pa$\cdot$s) such that the immersed length is $L=20$ cm. Spheres of different materials [Delrin (1.4 g/cm$^3$), borosilicate glass (2.4 g/cm$^3$) and stainless steel (8.05 g/cm$^3$)], with radii in the range 2--8 mm are released in close proximity to the top of the sheet. The  motion of the spheres is recorded at 30 frames/second with a camera (Nikon D5100). The displacements $y(t)$ and $h(t)$ of the sphere for a typical experiment are indicated as gray open circles in Fig.~\ref{FigSampleTraj}. The normal velocity depends strongly on the properties of both the sheet and the sphere, and is always smaller than the sedimentation velocity along the $\ey$ direction (see Fig.~\ref{FigSampleTraj}). We note that the Reynolds number $(\rho_f V_{\parallel} a/\mu)$ in all the experiments is smaller than $0.02$, so that inertial contributions to $V_{\perp}$ are at least an order of magnitude smaller than the observed migration velocities. 

Using a linear extension experiment, we measure the Young modulus of the sheets as $E \approx 245$ kPa, from which we calculate their bending rigidity $B = \frac{E b^3}{12 (1-\nu^2)}$, using $\nu \approx 0.48$ as the Poisson ratio for silicone rubber \citep{landau_lifschitz86_elasticity}.  
Since the sheets are suspended under their own weight, the tension varies in the direction of gravity.  
We approximate the tension by its mean, i.e. we use $T = \frac{1}{2}(\Delta \rho_s) g L b$, where $\Delta \rho_s = \rho_f - \rho_s$, in order to compare the experimental data with our theoretical predictions; relaxing this simplification yields only minor differences. 
Thus, on using \eqref{LambdaBetaH} and the above expressions for $B$ and $T$, we find $\tau= \frac{12(1-\nu^2)g L (\Delta \rho_s)}{E (b/a)^2}$, which is greater for thinner sheets and larger particles, and is in the range $0.25$--$8$ in our experiments. 
Recalling that the relative magnitude of tension and bending are determined by $\tau H$ [see e.g. \eqref{VperpGrav}], we expect bending to be important at small separation distances $\tau H \ll 1$, and for tension to dominate for larger separation distances with $\tau H \gg 1$. As we will show, our experimental data span both of these limiting regimes. 

To capture the observed trajectory, $y(t),\, h(t)$, we numerically integrate the velocity components, $V_{\perp}$ \eqref{VperpGrav} and $V_{\parallel}$. Figure \ref{FigSampleTraj} shows the trajectory of an experiment with a particular combination of particle size and density, and sheet bending rigidity. The theoretical predictions for the displacement, without adjustable parameters, are plotted as solid lines in Figs. \ref{FigSampleTraj}(a) and (b), showing that in order to achieve quantitative agreement with measured trajectories (gray circles) both bending (dashed) and tension (dotted) contributions must be incorporated. 

The system undergoes a transition between bending and tension dominated regimes as the separation distance increases over time. This transition is indicated for a single trajectory (black circles in Fig.~\ref{FigF}), but is clearly observed when varying system parameters such as particle size and density as well as the sheet bending rigidity, as shown in Fig.~\ref{FigF}. 
Neglecting the weak (logarithmic) dependence of $V_{\parallel}$ on $H$, the normal velocity is constant in the bending dominated regime ($H \ll \tau^{-1}$), i.e. the separation increases linearly with time, $h \approx \frac{3 \mu a^2 V_{\parallel}^2}{25 B} t$ (see Eq.~\ref{Vperp}). In our experiments, we do not observe trajectories that are characterized by bending alone, although bending is likely to dominate for smaller particles (since $\tau \propto a^{2}$). 

\begin{figure}[h!]
\centering
\includegraphics[scale=1]{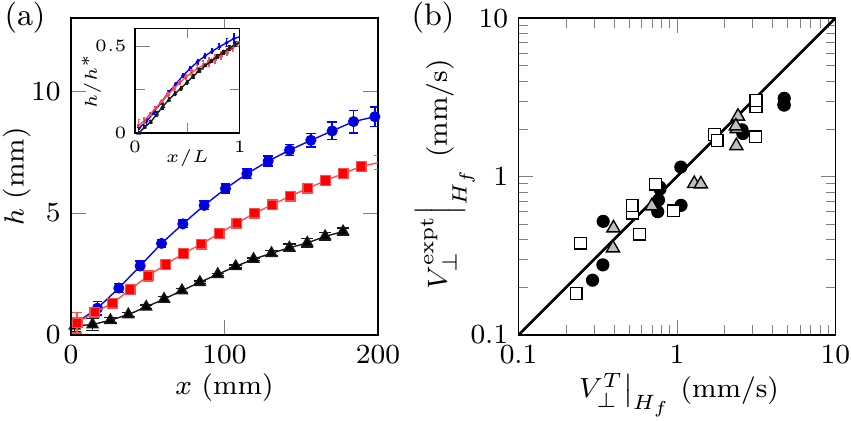}
\caption{(a) Experimental trajectories of three different spheres (circles: glass, $a_p =5$ mm; squares: glass, $a_p =3$ mm; triangles: Delrin, $a_p = 6$ mm) near a sheet with thickness $b=0.25$ mm, showing that the normal migration depends on both the size and the density of the particles. Inset shows rescaled trajectories for tension-dominated dynamics as explained in the text. (b) By the end of their trajectories ($H = H_f$), the normal velocity is dominated by tension; measured velocities (symbols) are plotted against the tension-dominated theoretical prediction, $V_{\perp}^{T}\big|_{H_f} = \frac{4 a^4 g^2 (\Delta \rho_p)^2}{2025 \mu T H_f \mathcal{R}_{\parallel}^2(H_f)}$. The data spans all sheet thicknesses [circles, $b=$ 0.25 mm; squares, $b=$ 0.38 mm; triangles, $b=$ 0.5 mm] and sphere properties (not indicated).} 
\label{FigTensionDominated}
\end{figure}

In the tension dominated limit ($\tau H \ll 1$; long times or thin sheets), integrating \eqref{Vperp} yields $h \approx \sqrt{2\mu a t/(25 T)} V_{\parallel}$, again neglecting the dependence of $V_{\parallel}$ on $H$. For our gravity-driven experiments, this expression reduces to $h \approx  h^*\sqrt{x/L}$, where $h^* = \frac{2 \sqrt{2}}{15}\left(\frac{\Delta \rho_p}{\Delta \rho_s}\right)^{\frac{1}{2}} a^{\frac{3}{2}} b^{-\frac{1}{2}}$. We find a good collapse of the experimental trajectory data for different particle sizes, as indicated in Fig. \ref{FigTensionDominated}(a).  For all the experiments, we can extract data points where the system is tension dominated. In particular, the measured $V_{\perp}$ of each particle by the end of its trajectory (for which particles assume final separation distances $H = H_f$) is well described by the tension-dominated limit of the theory, $V_{\perp}^{T}|_{H_f} = \frac{4 a^4 g^2 (\Delta \rho_p)^2}{2025 \mu T} \!\times\! \frac{1}{H_f \mathcal{R}_{\parallel}(H_f)}$ , as shown in Fig. \ref{FigTensionDominated}(b).

We comment on extensions of the theoretical framework developed here. 
It is possible to include a confining potential $G$ (units of force per volume), such that the membrane deformation is governed by $p^{(0)} = - (B \nabla^4 + T \nabla^2 + G) \zeta$. For biological membranes, a confining potential can arise from a nonzero curvature \citep{sei1997}, finite system size, or an underlying cytoskeleton \citep{fou04_fluctuation, bic06_brownian}; such a potential arises in macroscopic systems due to a body force (often gravity) acting normal to the membrane. 
The effect of this added potential to the normal velocity $V_{\perp}$ is accounted for by modifying the denominator of (\ref{Vperp}b) to $(k^4 + \tau H k^2 + \gamma H^2)$, where $\gamma = G \ell^4/(B H^2) = 4 G a^4/B$. For a cylindrical object translating perpendicular to its axis [here, the $x$ axis, cf. Fig. \ref{FigSetup}(b)], the only modification to $V_{\perp}$ involves replacing $\rmK_0(k)$ by $\frac{5}{3}  \sqrt{\frac{2}{k}} \rme^{-k}$ in (\ref{Vperp}); the results of \citet{sek93} and \citet{sko04} correspond to setting the denominator to a constant.

We have demonstrated theoretically and experimentally that a particle moving along a membrane will experience a hydroelastic repulsion. There are several specific consequences of this effect:
(1) A bacteria ($\sim 1\, \mu$m) swimming near a membrane ($B \sim 10$ $kT$) at its nominal speed ($\sim 30 \,\mu$m/s) will be hydro-elastically repelled at a speed comparable to its own. 
Both swimming and non-swimming bacteria, bacteriophage or sperm cells approaching a bio-membrane are expected to experience such repulsion.
(2) Our model suggests the possibility of a non-intrusive measurement of the elastic properties of biological membranes, one that does not require thermal equilibrium (e.g. by using a micron-sized bead and optical or magnetic tweezers \cite{boa14}).
(3) The lift force acting on a microscopic swimmer propelled in proximity to a bio-membrane is expected to be four orders of magnitude greater than the lift generated by a compressible surface (such as the cytoskeleton $G \sim 1000$ Pa) in the soft lubrication limit  \cite{sko04, sko05, sai16}. On the macroscopic scale, the particles in our experiments experienced a force greater by three orders of magnitude than a comparable experiment with a soft substrate.
(4) Small thermal fluctuations of the bio-membrane result in a reduced effective bending rigidity \cite{bro1975, fou04_fluctuation, bic06_brownian} and are therefore expected to enhance the repulsive migration due to hydro-elastic surfing. (5) Our macroscopic sedimentation experiments show that the hydro-elastic repulsion is sensitive enough to the size of the particle for sorting and separation purposes.

\begin{acknowledgments}
The authors acknowledge partial support from the Carbon Mitigation Initiative of Princeton University. We thank T. Salez for preliminary discussions, M. J. Shelley for helpful ideas and J. Nunes, A. Perazzo and  Y. E. Yu for assistance with the experiments. 

B. Rallabandi and N. Oppenheimer contributed equally to this work. 
\end{acknowledgments}

\bibliographystyle{apsrev4-1}
%\bibliography{SoftLub} 
\bibliography{particleSheet}
%,particleforces}

\end{document}